\documentstyle[prl,aps,floats,epsf]{revtex}
\begin{document}
\twocolumn[
\hsize\textwidth\columnwidth\hsize\csname@twocolumnfalse\endcsname
\draft
\title{Optical evidence for mass enhancement of quasiparticles in pyrochlore Cd$_2$Re$_2$O$_7$}
\author{N. L. Wang$^{1,2}$, J. J. McGuire$^{2}$, and T. Timusk$^{2}$}
\address{$^{1}$Institute of Physics, Chinese Academy of Sciences, P.O.Box 2711, Beijing 100080, P. R. China}

\address{$^{2}$Department of Physics and Astronomy, McMaster University, Hamilton, Ontario L8S 4M1, Canada}

\author{R. Jin, J. He, and D Mandrus}
\address{Solid State Division, Oak Ridge National laboratory, Oak Ridge, TN 37831, USA}
\maketitle
\begin{abstract}
We report on the results of optical studies of the newly
discovered superconductor Cd$_2$Re$_2$O$_7$ in the normal state.
We show that the compound has an exotic metallic state at low
temperature. The optical conductivity spectrum exhibits two
distinct features: a sharp renormalized resonance mode at zero
frequency and a broad mid-infrared excitation band. Detailed
analysis reveals a moderate enhancement of the effective mass at
low temperature and low frequency.

\end{abstract}

]
\epsfclipon

The magnetically frustrated pyrochlore oxides, which have a general formula
A$_2$B$_2$O$_7$ where B stands for a transition metal, have attracted
considerable interest recently. The B cations are six-fold coordinated and
located within distorted octahedra. Those octahedra are corner-sharing and form
a three-dimensional network.\cite{Subramanian} Many of the compounds undergo a
metal to non-metal transition without an associated structural change.
Cd$_2$Re$_2$O$_7$ is one of a few exceptions, which displays solely metallic
behavior below room temperature. This compound undergoes a second-order phase
transition at around 200 K and enters a better metallic state in low
temperature.\cite{Jin} Very recently it was found that Cd$_2$Re$_2$O$_7$
becomes superconducting below 2 K. This makes this compound the first
superconductor in the pyrochlore family.\cite{Hanawa,Sakai,Jin2} It is of great
interest to investigate the peculiar electronic state lying behind the
superconductivity. For this purpose we have investigated the optical properties
of the compound at different temperatures in the normal state. An exotic
metallic phase with strongly correlated electrons was revealed in the study.

The single crystals of Cd$_2$Re$_2$O$_7$ were grown using a vapor-transport
method described in detail in ref \cite{He}. They were well characterized by
x-ray diffraction, electron diffraction, resistivity, specific heat and
magnetic susceptibility measurements, showing superconductivity below T$_c
\sim$ 1.5 K.\cite{Jin,Jin2,He} The crystal structure is face-centered cubic.
Near normal incidence reflectivity spectra, R($\omega$), were measured from 30
to 40000 cm$^{-1}$ on a natural growth surface normal to the a-axis. Standard
Kramers-Kronig transformations were employed to derive the frequency-dependent
conductivity and dielectric function.

Cd$_2$Re$_2$O$_7$ exhibits an unusual temperature-dependent dc
resistivity in the normal-state: it is almost T-independent at
high temperature but decreases rapidly below 200 K. The behavior,
displayed in the inset of Fig. 1, is inconsistent with the
traditional electron-phonon scattering mechanism, which should
yield a linear T-dependent behavior at high temperature. Below 50
K, the dc resistivity follows an approximately quadratic
T-dependence, implying electron-electron scattering and a
Fermi-liquid-like state at low temperature. The reflectivity data
at 300 K, 150 K and 24 K are shown in Fig. 1. We note immediately
that the reflectivity in the far-infrared spectral range increases
with decreasing temperature, characteristic of metallic response.
However, the reflectivity in the mid-infrared range decreases with
decreasing temperature. This suggests a redistribution of spectral
weight with decreasing temperature, which should be seen more
clearly in the frequency-dependent conductivity spectra. The
reflectivity at high frequency is $T$-independent. A plasma edge
minimum is seen at frequency close to 15000 cm$^{-1}$.

The real part of the conductivity, $\sigma_1(\omega)$, is shown in
Fig. 2. We use the Hagen-Rubens relation for the low frequency
extrapolation in the Kramers-Kronig analysis. The conductivity
spectrum at room temperature exhibits a number of phonon modes
(170 cm$^{-1}$, 280 cm$^{-1}$, 372 cm$^{-1}$, 570 cm$^{-1}$)
together with broad electronic excitations.\cite{Note1} As
temperature decreases, $\sigma_1(\omega)$ is significantly
enhanced in the far-infrared range through spectral weight shifted
from mid-infrared electronic excitations below 3300 cm$^{-1}$.
From the spectrum at 24 K, one can identify two distinct features:
a narrow Drude-like resonance at $\omega$=0 and very broad
mid-infrared excitations. We emphasize here that the Drude-like
peak is not a consequence of the low-frequency extrapolation since
it is found that different extrapolations almost do not affect the
spectra in the measured frequency range. The conductivity spectra
differ markedly from the optical response for a simple metal. The
observation highlights the many-body nature in the coherent
metallic state of Cd$_2$Re$_2$O$_7$ at low temperature. We noticed
that the conductivity spectra resemble those of heavy-Fermion (HF)
systems where similar and even narrower resonances are found in
the low frequency conductivity in nearly all HF metals. In
addition to the effect induced by the electron-electron
correlation, the electron-phonon interaction seems to be strong at
low temperature as well. Except for the phonon mode at 570
cm$^{-1}$, almost all other phonon modes shift to higher
frequencies (hardening) in low temperature metallic state. A new
phonon mode appears at 340 cm$^{-1}$  at 150 K and 24 K. This is
consistent with the electron diffraction measurement which
indicated a structural modulation associated with the phase
transition at 200 K.\cite{Jin}

\begin{figure}[t]
\centerline{\epsfxsize=2.6 in \epsffile{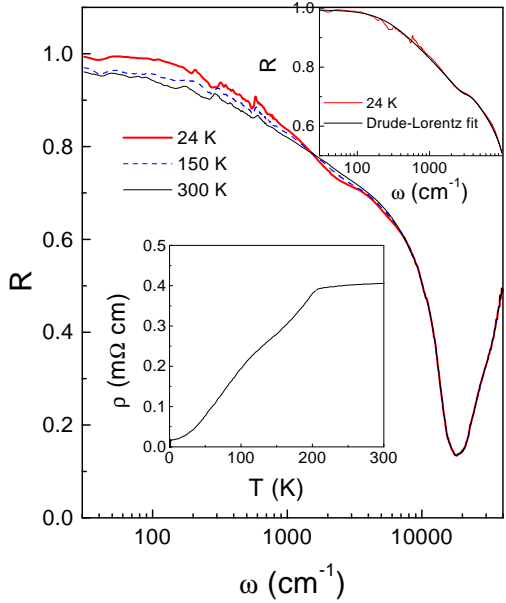}}
\vspace{0.1in} \caption{The frequency dependent reflectivity of
Cd$_2$Re$_2$O$_7$ at 300 K, 150 K and 24 K. The lower inset shows
the dc resistivity as a function of temperature. The upper inset
shows the Drude-Lorentz fit to the low-T reflectivity curve over a
broad frequency range.} \label{1}
\end{figure}
\begin{figure}[t]
\centerline{\epsfxsize=2.8 in\epsffile{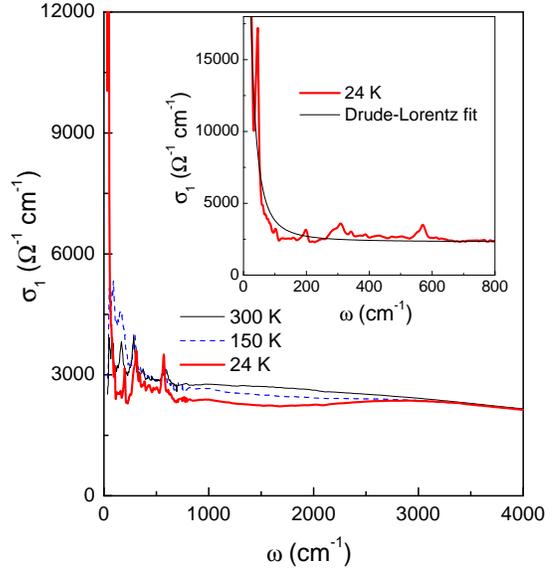}}
\vspace{0.1in} \caption{The frequency dependent conductivity of
Cd$_2$Re$_2$O$_7$ at 300 K, 150 K and 24 K. Inset shows the
Drude-Lorentz fit to the low-$\omega$ conductivity spectrum at 24
K.} \label{1}
\end{figure}

To isolate the different components of the electronic excitations, we
fit simultaneously {\em R($\omega$)} and $\sigma_1(\omega)$ at 24
K to a sum of a Drude term and Lorentz oscillators. This will
reduce the uncertainty caused by the extrapolation in
Kramers-Kronig analysis. The general formula for the complex
dielectric function is\cite{Timusk}

\begin{equation}
\overline{\epsilon}(\omega)=
\epsilon_\infty - {\omega_p^{*2} \over \omega^2 + i\omega\Gamma_D}
+ \sum{\omega_{p,j}^2 \over (\omega_j^2-\omega^2) - i\omega\Gamma_j}.
\label{chik}
\end{equation}
where $\epsilon_\infty$ is the dielectric constant at high
frequency, $\omega_p^*$ and $\Gamma_D$ in the Drude term are the
plasma frequency and the relaxation rate of the free charge
carriers, while $\omega_j$, $\Gamma_j$, and $\omega_{p,j}$ are the
resonance frequency, the damping, and the mode strength of the
Lorentz oscillators, respectively. As shown in the inset of fig. 1
and fig.2, both the {\em R($\omega$)} and $\sigma_1(\omega)$ can
be well fit to a Drude part and two Lorentz oscillators over a
broad frequency range.\cite{Note2} The Drude component has a
plasma frequency of $\omega_p^*$=6800 cm$^{-1}$ and width 21
cm$^{-1}$. The Lorentz parameters are $\omega_{p,1}$=35350
cm$^{-1}$, $\Gamma_1$=9000 cm$^{-1}$, $\omega_1$=400 cm$^{-1}$,
and $\omega_{p,2}$=5500 cm$^{-1}$, $\Gamma_2$=1910 cm$^{-1}$,
$\omega_2$=3270 cm$^{-1}$. We can see that the Drude part, though
dominating the static conductivity, has only a fraction of the
total spectral weight below the frequency of reflectivity minimum.

From the fitting parameters of the Drude component, the estimated
conductivity at static limit, in terms of the simple Drude model,
is around 38500 $\Omega^{-1}cm^{-1}$. This is in agreement with
the reported dc resistitivity data.\cite{Hanawa} The dc
experimental data at 24 K in our measurement (see the inset of
fig. 1) is about 34000 $\Omega^{-1}cm^{-1}$, which is a bit
smaller than the zero-frequency extrapolation of the optical data.
However, such discrepancy is within the systematic errors of
experiments. Since the reflectance at far-infrared region is close
to unit, the uncertainty of the reflectance even within 0.5$\%$
could give substantial difference of conductivity in static limit.
In addition, the uncertainty of the geometric factors also adds
some errors to the dc resistivity data .

An alternative way of making quantitative comparisons between the
low-frequency resonance mode and overall spectral weight below the
interband transition is to sum the spectral weight under the
conductivity spectrum. The unscreened optical plasma frequency can
be estimated by summing the spectral weight below the frequency of
interband transition,
$8\int_0^{\omega^\prime}\sigma_1(\omega)d\omega = {4\pi
n{e^2}/m_B} = \omega_p^2$, where $\omega^\prime$ is a cutoff
frequency. By integrating the conductivity to 15000 cm$^{-1}$
where the conductivity spectrum shows a well-defined minimum, we
get $\omega_p$=3 $\times$ 10$^4$ cm$^{-1}$. This value is
approximately the same for calculations at the three different
temperatures. This plasma mode is clearly seen in the optical
reflectivity as the high frequency plasma edge and in the real
part of dielectric function as a zero-crossing at high frequency.
Similarly, by applying sum-rule arguments, we can calculate
the spectral weight below the narrow Drude-like mode and obtain
another plasma frequency as $\omega_p^{*2}$ = 4$\pi n{e^2}/m^*$ =
8$\int_0^{\omega_c}\sigma_1(\omega)d\omega$, where m$^*$ is the
effective mass at low frequency and $\omega_c$ is another cutoff
frequency only for the narrow Drude-like mode. The value of
$\omega_p^*$ we obtained is 6720 cm$^{-1}$. This value is close to
the one we obtained from the above Drude-Lorentz fit, but is
significantly smaller than the unscreened plasma frequency.

\begin{figure}[t]
\centerline{\epsfxsize=2.7 in \epsffile{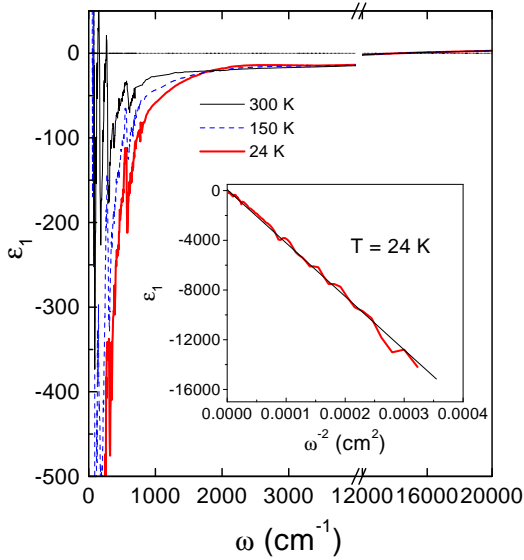}}
\vspace{0.1in} \caption{The frequency dependence of the dielectric
functions of Cd$_2$Re$_2$O$_7$ at 300 K, 150 K and 24 K. The
high-frequency plasma mode is seen as the zero-crossing feature at
$\omega$ $\approx$ 15000 cm$^{-1}$. Inset shows the $\epsilon_1$
vs $\omega^{-2}$ plot at low temperature.} \label{1}
\end{figure}

The occurrence of a sharp and narrow resonance mode at $\omega$=0
together with very broad mid-infrared excitations in the optical
conductivity of strongly correlated systems was widely explained
in terms of the renormalized quasi-particles in the many-body
picture.\cite{allen77,Bonn,Degiorgi1} It was suggested that the
finite energy modes, basically containing all the spectral weight
at high temperature, are associated with the unrenormalized band
mass m$_B$. As the temperature is lowered, correlation effects
dressing the free charge carriers are manifested through the
redistribution of the spectral weight between the higher and lower
frequency. The sharp and narrow Drude-like component appears as a
consequence of the enhancement of both the effective mass and the
scattering time. With the assumption that the total charge-carrier
density does not change with temperature, it is possible to
directly estimate the enhancement of the effective mass by
m$^*$/$m_B$ = {$\omega_p^2$/$\omega_p^{*2}$}, which gives value of
20. Indeed, specific heat measurements on Cd$_2$Re$_2$O$_7$
indicated an enhanced electronic specific heat coefficient,
$\gamma$ $\sim$ 30 mJ/mol-K$^2$.\cite{Jin2,Hanawa} The behavior of
the upper critical field H$_{c2}$ vs T also indicates that the
Cooper pairs are composed of rather heavy quasiparticles.
Qualitatively, the optical measurement is in agreement with the
enhancement picture of effective mass.

It should be noted that the enhancement factor m$^*$/$m_B$ refers to the
enhanced effective mass in relation to the band mass, which could be
significantly larger than the free-electron mass $m_e$.\cite{Degiorgi1,Awasthi}
According to the available band structure calculation of this
material,\cite{Singh} the Fermi surfaces consist of nearly spherical electron
pockets centered at $\Gamma$ point and very heavy hole section at the zone
boundary. The calculated density of states near E$_F$ is derived mainly from
the heavy hole bands, which produces the bare band specific heat coefficient as
large as $\gamma$=12.4 mJ/mol K$^2$. Comparing with the measured value of 30
mJ/mol K$^2$, we obtain a mass enhancement due to many-body effect
m$^*$/$m_B$=2.4. However, this value is much smaller than the value of 20
obtained from optical conductivity analysis which is solely due to the electron
correlation effects. One possible explanation for the contradiction is that the
mass enhancements in specific heat and transport are associated with different
bands in the electronic structure. The mass enhancement in specific heat mainly
comes from the heavy hole sheets with small correlation effect, on the other
hand, the transport is dominated by the light electron band with strong
renormalization effect by electron correlations. The later argument is also
supported by Hall effect measurement.\cite{Jin2}

Fig. 3 shows the real part of the dielectric function,
$\epsilon_1(\omega)$, as a function of frequency. The
zero-crossing frequency at around 15000 cm$^{-1}$, corresponding
to the plasma edge in the reflectance spectrum, represents the
screened overall plasma frequency. The contribution of the broad
mid-infrared bands to the real part of the dielectric function,
$\epsilon_1(\omega)$, varies very slowly with $\omega$ over a
large frequency range below the frequency of zero-crossing. The
low-frequency $\epsilon_1(\omega)$ at high temperature can become
even positive because of the contributions from phonon modes and
some localized effect of quasiparticles.

\begin{figure}[t]
\centerline{\epsfxsize=2.6 in \epsffile{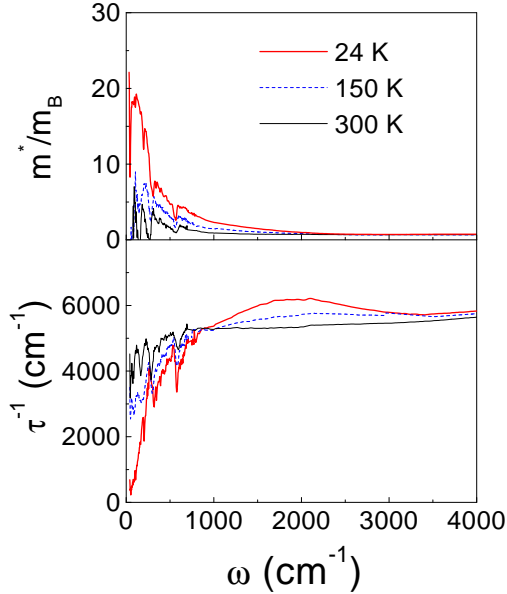}}
\vspace{0.1in} \caption{The frequency dependence of the
quasiparticle effective mass and scattering rate of
Cd$_2$Re$_2$O$_7$ at 300 K, 150 K and 24 K.} \label{1}
\end{figure}

Let us analyze the low-T spectrum of $\epsilon_1(\omega)$ in the
low frequency range due to the contribution from the renormalized
Drude-like component. Because the contribution of the broad
mid-infrared bands to $\epsilon_1(\omega)$ varies slowly in
$\omega$, in the case of $\omega$ $\gg$ $\Gamma_D$ the real part
of dielectric function, in terms of equation (1), can be
approximated as $\epsilon_1(\omega) \approx \epsilon_x -
{\omega_p^{*2}/\omega^2}$, where $\epsilon_x$ represents a
background dielectric constant at a high frequency determined by
$\epsilon_\infty$ and contributions from broad mid-infrared
excitations. Then, the slope in a $\epsilon_1$ vs $\omega^{-2}$
plot will provide the value of the $\omega_p^{*2}$ for the
renormalized Drude-like component. The inset of Fig.3 shows the
$\epsilon_1$ vs $\omega^{-2}$ plot. The solid and dash lines are
the experimental and the linear fit to the data. The value of
$\omega_p^*$ extracted is 6620 cm$^{-1}$, which is in good
agreement with those determined in the above analysis.

The frequency-dependence of the scattering rate as well as of the
effective mass can be alternatively determined by using the
generalized Drude model by
\begin{equation}
{m^* \over m_B} = {\omega_p^2 \over 4\pi\omega} {\sigma_2(\omega) \over
(\sigma_1^2(\omega) + \sigma_2^2\omega)} \label{dclim}
\end{equation}
\begin{equation}
{1 \over \tau(\omega)} = {\omega_p^2 \over 4\pi} {\sigma_1(\omega)
\over \sigma_1^2(\omega) + \sigma_2^2\omega)},
\label{dclim}
\end{equation}
where $\omega_p^2$ is the unscreened overall plasma frequency.
This analysis is also quite often used to quantify the
renormalization effect of electronic correlation in HF
materials.\cite{Degiorgi1,Degiorgi3,Dordevic} The derived spectra
of the effective mass and of the scattering rate with $\omega_p$=3
$\times$ 10$^4$ cm$^{-1}$ are plotted in Fig.4. At room T, the
spectra of 1/$\tau$ depends weakly on $\omega$. As T lowered, the
1/$\tau$ is suppressed at low-$\omega$. In accord with this, the
effective mass is enhanced. We observed m$^*$/m$_B$ $\approx$ 20
for T=24 K, which is well in agreement with the value obtained
above.

Our study revealed unambiguously the renormalization effect by
strong electron correlation in Cd$_2$Re$_2$O$_7$, which makes the
optical response quite similar to other HF materials. The major
difference is that here the effective mass is relatively smaller
than typical HF metals with 4f and 5f electrons. This is
understandable considering the fact that the heavy electrons in
correlated system originate, in general, from the interaction of
the localized f or d electrons with the bands of delocalized
electrons. This leads to the strong mixing, or hybridization, of
the Fermi electrons with the localized f or d electrons. The final
result is a renormalization of the Fermi surface and a drastic
enhancement of the effective mass of free band electrons at E$_F$.
Cd$_2$Re$_2$O$_7$ is a transition-metal system with 5d electrons.
The 5d wave function is much more spatially extended than 4f or 5f
electrons, which naturally results in a relatively smaller
m$^*$/m$_B$. We speculate that, similar to other heavy Fermion
superconducting materials, the superconductivity observed in this
material is due to the condensate of those relatively heavy
electrons.

Finally we remark that Cd$_2$Re$_2$O$_7$ is not the unique oxide
material exhibiting an enhancement of the effective mass at low
temperature. The metallic mixed-valent compound LiVO$_4$, which
also has the pyrochlore lattice of transition metal V, is another
example showing unusual and strong HF behavior at low T. It should
be noted that, in such oxide materials, both the localized
electrons and delocalized band electrons should originate from the
d electrons of the transition metals. At present, the physical
origin of the heavy quasiparticle excitations is far from clear.
It is suggested that it differs considerably from that of other
known HF systems.\cite{Fulde,Laad}

In conclusion, our optical study reveals that the low-T metallic
state of Cd$_2$Re$_2$O$_7$ is quite different from a simple metal.
The spectra consist of two distinct charge excitations: a sharp
renormalized resonance mode at zero frequency and broad
mid-infrared excitation. The analysis reveals a moderate
enhancement of the effective mass at low temperature and
frequency. We compare the optical effective mass enhancement with
that from thermodynamical data. Our analysis suggests that
different bands in the electronic structure are associated with
the enhancements of specific heat and transport.

Wang was in part supported by National Science Foundation of China
(No.10025418). The work at McMaster University was supported by
the Natural Sciences and Engineering Research Council of Canada.
Oak Ridge National laboratory is managed by UT-Battelle, LLC, for
the U.S. Department of Energy under contract DE--Ac05-00OR22725.

\end{document}